\documentstyle[12pt]{article}
\def\lsim{\mathrel{\rlap{\raise 2.5pt \hbox{$<$}}\lower 2.5pt}}
\def\gsim{\mathrel{\rlap{\raise 2.5pt \hbox{$>$}}\lower 2.5pt}}
\begin{document}
\bibliographystyle{plain}
\thispagestyle{empty}
\vspace{-5mm}
\begin{center}
{\bf  Infra-red stable fixed points of Yukawa couplings in
non-minimal supersymmetric standard model with R-parity violation}\\
\vspace{1.0cm}
P. N. Pandita and P. Francis Paulraj\\
\begin{small}
Department of Physics, North-Eastern Hill University,\\
Shillong 793 022, India
\end{small}
\end{center}
\vspace{1cm}
\begin{abstract}
We study the renormalization group evolution and the 
infra-red stable fixed points of the Yukawa couplings
of the non-minimal  supersymmetric standard model 
(NMSSM) with R-parity violation.
Retaining only the  R-parity violating couplings of  higher
generations, we analytically study the solutions of 
the renormalization group
equations of all the couplings of the model.
We find that there are no simultaneous non-trivial infra-red fixed 
points for all the couplings of the model,
and that the infra-red fixed point structure of the model
is similar to the MSSM with R-parity violation.
In particular, we show that
only the baryon number violating coupling $\lambda^{''}_{233}$, 
together with top- and bottom-quark
Yukawa couplings, approaches  a non-trivial
infra-red stable fixed point. 
However, this fixed point solution predicts 
a top-quark Yukawa coupling which is incompatible with the 
top quark mass for any value of $\tan\beta$.
\end{abstract}
\vspace{0.5cm}
{\it PACS No.:  11.10.Hi, 11.30.Fs, 12.60.Jv}\\ 
{\it Keywords:  Supersymmetry, R-parity violation, Infra-red fixed points}
\vspace{1.0cm}
\begin{center}
(To appear in Phys. Lett. B)
\vspace{1.0cm}
\end{center}

\newpage

Recently considerable  attention has been devoted to the study of
infra-red~(IR)
stable fixed points~\cite{pendross} of the  Yukawa
couplings of the minimal supersymmetric standard model~\cite{schrempp}.
In the minimal supersymmetric standard model(MSSM)~\cite{nilles}, 
gauge invariance and supersymmetry allow dimension four 
couplings which violate~\cite{weinberg} baryon ($B$) and lepton number
($L$).  These couplings are usually forbidden by invoking a 
discrete symmetry~\cite{farrar} called R-parity ($R_p$). 
However, the assmption of
$R_p$ conservation at the level of MSSM appears to be {\it ad hoc}, since
it is not required for the internal consistency of the model. 
Much attention has, therefore, been paid to the study of
the renormalization group (RG) evolution of these 
$B$- and $L$-violating (and $R_p$ violating) Yukawa
couplings. This includes the study of the quasi-fixed point 
behaviour~\cite{barger}, as well as the true 
infra-red fixed points of the different Yukawa couplings and 
the analysis of their stability~\cite{anant}.
This has led to certain insights and constraints on the fixed
point behavior of some of the $R_p$ violating Yukawa couplings, involving
higher generation indices.   
For the fixed point or quasi-fixed point scenarios
to be successful, it is necessary that
these fixed points be stable~\cite{allanach}.
It has been shown that in the minimal supersymmetric standard model
with R-parity violation, only the $B$-violating coupling 
$\lambda''_{233}$, together with top- and bottom-quark Yukawa couplings,
approaches a non-trivial  infra-red stable fixed point, whereas 
all other non-trivial fixed point solutions are either unphysical
or unstable in the infra-red region~\cite{anant}.

It is well known that the minimal supersymmetric standard model 
suffers from the so-called $\mu$ problem associated with the 
bilinear term connecting the two-Higgs doublet superfields
in the superpotential. A simple solution to this problem is
to postulate the existence of a Higgs singlet superfield,
and to couple it to the two Higgs doublets in the superpotential
via a dimensionless trilinear coupling. When the Higgs singlet obtains 
a vacuum expectation value, a bilinear term involving the two Higgs
doublets is naturally generated\cite{fayet}. However, this leads to 
additional trilinear superpotential couplings in the model,
the so called non-minimal supersymmetric standard model(NMSSM).
Because of the additional trilinear couplings, it is
important to study the infra-red fixed points of the NMSSM,
and analyze the effect of these additional trilinear
Yukawa couplings on the infra-red behaviour  the other Yukawa and 
$B$- and $L$-violating couplings, and contrast it with the 
situation that obtains in the MSSM.

To this end we shall consider renormalization
group evolution and infra-red fixed points
of the Yukawa couplings of the non-minimal supersymmetric
standard model (NMSSM) 
with baryon and lepton number (and $R_p$) violation.
The superpotential of the NMSSM is given by 
\begin{equation}\label{nmssmw}
W = (h_U)_{ab} Q^a_L \overline{U}^b_R H_2 
+ (h_D)_{ab} Q^a_L \overline{D}^b_R H_1 + (h_E)_{ab} L^a_L 
\overline{E}^b_R H_1 + \lambda N H_1H_2 - \frac{k}{3}N^3.
 \end{equation}
Gauge invariance, supersymmetry and renormalizability allow the 
following superpotential $B$ and $L$ violating terms in the NMSSM:
\begin{eqnarray}
& \displaystyle W_B={1\over 2}\lambda''_{abc} \overline{D}^a_R
\overline{D}^b_R \overline{U}^c_R,  & \label{Bviolating} \\
& \displaystyle W_L= \lambda_{a} NL_a H_2  
+ {1\over 2}\lambda_{abc} L^a_L L^b_L \overline{E}^c_R
+ \lambda'_{abc} L^a_L Q^b_L\overline{D}^c_R, & \label{Lviolating}
\end{eqnarray}
where the notation~\cite{anant} is standard, except that there is 
a new  $L$-violating term with the dimensionless
Yukawa coupling $\lambda_a$ in
(\ref{Lviolating}). This term can be rotated away into the R-parity
conserving term $\lambda N H_1H_2$ via an $SU(4)$ rotation
between the superfields $H_1$ and $L_i$. However, this rotation 
must be performed at some energy scale, and the term is
regenerated through the renormalization group equations.
The Yukawa couplings
$\lambda_{abc}$ and $\lambda''_{abc}$ are antisymmetric in their
first two indices due to $SU(2)_L$ and $SU(3)_C$ group
structure, respectively.  Phenomenological
studies of supersymmetric models of this type
have placed  constraints~\cite{dreiner} on the various couplings
$\lambda_{abc}$, $\lambda'_{abc}$ and $\lambda''_{abc}$ in 
the MSSM, but there
is still considerable room left.  We note that the simultaneous presence
of the terms in Eq.~(\ref{Bviolating}) and 
Eq.~(\ref{Lviolating}) 
is essentially ruled out by the stringent constraints~\cite{smirnov}
implied by the lack of
observation of nucleon decay.

In addition to the dominant third generation Yukawa couplings
$h_t \equiv (h_U)_{33}$, $h_b \equiv (h_D)_{33}$ and $h_{\tau} \equiv
(h_E)_{33}$, and the trilinear couplings $\lambda$ and $k$
in the superpotential~(\ref{nmssmw}), there are 9 independent 
$R_p$ violating couplings
$\lambda''_{abc}$ in Eq.~(\ref{Bviolating}), and 39 independent
$\lambda_{abc}$,  $\lambda'_{abc}$ and $\lambda_a$
in Eq.~(\ref{Lviolating}).
Thus, one would have to solve 14
coupled non-linear evolution equations in the B-violating case,
and 44 in the L-violating case, in order to study 
the evolution of the Yukawa couplings in the 
nonminimal supersymmetric standard model with $R_p$ violation.
In order to render the Yukawa coupling evolution tractable, we
need to make certain plausible 
simplifications~\cite{barger}~\cite{anant}.  
As is usually done in the analysis of the MSSM with R-parity
violation, we shall
retain only the highest generation couplings $\lambda_3$,
$\lambda_{233}$, $\lambda'_{333}$ and $\lambda''_{233}$, 
and neglect the rest.  We note that
the $R_p$ violating couplings to higher generations evolve more
strongly because of larger Higgs couplings in their evolution equations,
and hence could take larger values than the corresponding couplings to
the lighter generations.  Furthermore, the experimental upper limits
are stronger for the $R_p$ violating Yukawa couplings corresponding
to the lighter generations.

We shall first consider the RG evolution of Yukawa couplings
arising from superpotential terms 
in~(\ref{nmssmw}) and (\ref{Bviolating}),
which involve baryon number violation.
We have derived the one-loop renormalization group 
equations for all the trilinear couplings of the model.
For the couplings $h_t,\, h_b, \,
h_\tau, \, \lambda, \, k$ and $\lambda''_{233}$ 
(all other Yukawa couplings  set to zero) these can be written as:
\begin{eqnarray} 
& \displaystyle 16 \pi^2 {dh_t\over d(\ln\, \mu)}=
h_t\left(6 h_t^2 + h_b^2 + \lambda^2 + 2\lambda''^2_{233}-{16\over 3} g_3^2
-3 g_2^2 -{13\over 15} g_1^2\right), & \nonumber \\
& \displaystyle 16 \pi^2 {dh_b\over d(\ln\, \mu)}=
h_b\left(  h_t^2 + 6h_b^2+h_\tau^2 + \lambda^2 
+ 2\lambda''^2_{233}-{16\over 3} g_3^2
-3 g_2^2 -{7\over 15} g_1^2\right), & \nonumber \\
& \displaystyle 16 \pi^2 {dh_\tau\over d(\ln\, \mu)}=
h_\tau\left(  3 h_b^2+4 h_\tau^2 + \lambda^2
-3 g_2^2 -{9\over 5} g_1^2\right), & \label{htauequation} \\
& \displaystyle 16 \pi^2 {d\lambda\over d(\ln\, \mu)}=
\lambda\left(3 h_t^2 +  3 h_b^2 +  h_\tau^2 + 4 \lambda^2
+ 2 k^2 - 3 g_2^2 -{3\over 5} g_1^2\right), & \nonumber\\
& \displaystyle 16 \pi^2 {dk\over d(\ln\, \mu)}=
6 k \left( \lambda^2 + k^2\right), & \nonumber\\
& \displaystyle 16 \pi^2 {d\lambda''_{233}\over d(\ln\, \mu)}=
\lambda''_{233}\left(2 h_t^2 +2h_b^2+6\lambda''^2_{233}-8 g_3^2
-{4\over 5} g_1^2\right). & \nonumber 
\end{eqnarray}
We note that evolution equations
for the gauge couplings $(g_i, i = 1, 2, 3)$
in NMSSM are identical to
those in the MSSM, since the additional Yukawa couplings do not play
a role at this order. With the definitions
\begin{equation}\label{redefinitions}
R''={\lambda''^2_{233}\over g_3^2}, \,\,\,
R_\tau={h_\tau^2\over g_3^2},  \,\,\,  
R_b={h_b^2\over g_3^2}, \,\,\,  R_t={h_t^2\over g_3^2}, \,\,\,
R_\lambda={\lambda^2\over g_3^2}, \,\,\, 
R_k={k^2\over g_3^2},  \,\,\, 
\tilde{\alpha}_3 = {g_3^2\over (16\pi^2)},
\end{equation}
ordering the ratios as $R_i = (R'', R_{\tau}, R_b, R_t, R_{\lambda},
R_k)$, and 
retaining only the $SU(3)_C$ gauge coupling, 
we can 
rewrite the renormalization group equations (\ref{htauequation})
in the form ($t = - ln\mu^2$)
\begin{equation}\label{componentequation}
{d R_i\over dt}=\tilde{\alpha}_3 R_i
\left[(r_i+b_3)-\sum_j S_{ij} R_j\right].
\end{equation}
where $r_i=\sum_R 2 \, C_R$,  $C_R$ is the QCD Casimir
for the various fields ($C_Q=C_{\overline{U}}=C_{\overline{D}}=4/3$), 
the sum is over the representation of
the three fields associated with the trilinear coupling that enters
$R_i$, and $S$ is a matrix whose value is fully specified by
the wavefunction anomalous dimensions. 
A fixed point, denoted as $R_i^*$,  is then reached
when the right hand side of Eq.~(\ref{componentequation}) is
0 for all $i$.  We, then,  see that there are two fixed point 
values for each coupling:
$R_i^*=0$, or the non-trivial fixed point solution 
\begin{equation}\label{ristarequation}
R_i^*=\sum_j (S^{-1})_{ij} (r_j+b_3).
\end{equation}
Since we shall consider the fixed points of the couplings $h_t$, $h_b$ 
$\lambda$, $k$  and
$\lambda_{233}^{''}$ only, we shall ignore the evolution equation
for $h_{\tau}$.
However,  the coupling $h_{\tau}$ does enter
the evolution  of
$h_b$, but it can be related to $h_b$ at the weak scale (which we 
take to be the top-quark mass),  since
\begin{equation}\label{2ndhtaurelation}
h_\tau(m_t)={\sqrt{2} m_\tau(m_t)\over \eta_\tau v \cos \beta}
={m_\tau(m_\tau)\over m_b(m_b)} {\eta_b \over \eta_\tau} h_b(m_t)
= 0.6 h_b(m_t),
\end{equation}
where $\eta_b$ gives the QCD or QED running~\cite{barger2} 
of the b-quark mass $m_b(\mu)$ between $\mu = m_b$ and $\mu = m_t$
(similarly for $\eta_{\tau}$), 
and $\tan\beta=<H_2^0>/<H_1^0>$, 
with   $v=(\sqrt{2} G_F)^{-1/2}=246$ GeV.
The anomalous dimension matrix $S$ can, then, be written as 
\begin{equation}\label{Smatrix}
S=\left(
\begin{array}{ c c c c c}
6 & 2 & 2 & 0 & 0 \\
2 & 6 + \eta & 1 & 1 & 0 \\
2 & 1 & 6 & 1 & 0 \\
0 & 3 + \eta & 3 & 4 & 2\\
0 & 0 & 0 & 6 & 6\\
\end{array}
\right),
\end{equation}
with  ordering $(R^{''}, R_b, R_t, R_{\lambda}, R_k)$,
where $\eta=h_\tau^2(m_t)/h_b^2(m_t)\simeq 0.4$ is the factor
coming from Eq.~(\ref{2ndhtaurelation}). We, therefore, obtain
the following fixed point solution for the ratios: 
\begin{equation}
(R''^*, R_b^*,  R_t^*, R_{\lambda}^*, R_k^*)  = ({5 \over 12},
{25\over{4(10 + \eta)}}, {{5 (5 + \eta)} \over {4(10 + \eta)}},
-{{115 + 24 \eta} \over {4(10 + \eta)}}, 
{{95 + 22 \eta} \over {4(10 + \eta)}}). 
\label{rstarsolution}
\end{equation}
Since  $R_{\lambda}^*$ is negative, this is 
not an acceptable fixed point solution. Thus, there is 
no infra-red fixed point solution for NMSSM with baryon 
number violation where all the trilinear 
superpotential couplings attain non-zero fixed point values.

We next try to find a fixed point solution with $R_b^*=0$, and
$R''^*, R_t^*, R_{\lambda}^*$ and $R_k^*$
being given by their non-zero solutions, 
which is relevant for the low values of the parameter
$\tan\beta.$ We 
need to consider the appropriate $4\times 4$ sub-matrix
of the matrix S in Eq.~(\ref{Smatrix}) to obtain the fixed point
solutions for $R''^*, R_t^*, R_{\lambda}^*$ and $R_k^*$ in this 
case.  This sub-matrix is given by
\begin{equation}\label{Submatrix}
S=\left(
\begin{array}{ c c c c }
6 & 2 & 0 & 0 \\
2 & 6 & 1 & 0 \\
0 & 3 & 4 & 2 \\
0 & 0 & 6 & 6\\
\end{array}
\right),
\end{equation}
with the fixed point solution
\begin{equation}
R_b^*=0, ~~~~~~~ 
(R''^*,  R_t^*, R_{\lambda}^*, R_k^*)
= ({{95}\over{138}}, {{10}\over{23}}, -{{38}\over{23}}, {{53}\over{46}}).
\label{2ndrstarsolution} 
\end{equation}
This is not a theoretically acceptable solution, 
as all the fixed point values are not positive.

We now try to find an infra-red fixed point solution 
with the trlinear coupling $R_{\lambda}^* = 0$.
Proceeding in the same manner as in the case of 
$R_b^* = 0$, we get the fixed point solution 
\begin{equation}
R_{\lambda}^* = 0, ~~~~~~ 
(R''^*,  R_b^*, R_t^*, R_k^*)
= ({{385+76\eta}\over{12(85+16\eta)}}, {{44}\over{3(85+16\eta)}}, 
{{10+2\eta}\over{85+16\eta}}, -{{1}\over{2}}).
\label{3rdrstarsolution} 
\end{equation}
Since $R_k^* < 0$ this fixed point solution must be rejected.
We can also try to find a fixed point solution with  
$R_k^* = 0$ with the result
\begin{equation}
R_k^* = 0, ~~~~~
(R''^*,  R_b^*, R_t^*, R_{\lambda}^*)
= ({{455+83\eta}\over{30(25+4\eta)}}, {{37}\over{6(25+4\eta)}}, 
{{51(5+\eta)}\over{30(25+4\eta)}}, -{{435+96\eta}\over{30(25+4\eta)}}), 
\label{4thrstarsolution} 
\end{equation}
which is not an acceptable solution either.
Finally, we try a fixed point solution with $R''^* = 0$, and 
all other Yukawa couplings attaining non-zero fixed point values. 
This is the case of NMSSM with R-parity conservation.
In this case we find the solution 
\begin{equation}
R''^* = 0, ~~~~~~
(R_b^*, R_t^*, R_{\lambda}^*, R_k^*)
= ({{5}\over{3(10+\eta)}}, {{5(5+\eta)}\over{3(10+\eta)}}, 
-{{105+23\eta}\over{3(10+\eta)}}, {{90+19\eta}\over{6(10+\eta)}}), 
\label{5thrstarsolution} 
\end{equation}
which must also be rejected. We have, thus, shown that there is 
no infra-red fixed fixed solution with one of the trilinear couplings
being zero, and all others attaining a nonzero fixed point value.

Having failed to find an acceptable fixed point solution with
one of the couplings being zero, we now try to find a
solution with two of the trilinear coupling attaining a zero
fixed point value. In this case we find only the following acceptable
fixed point solution: 
\begin{equation}
R_{\lambda}^* = R_k^* = 0, ~~~~~~~~
(R''^*,~ R_b^*,~ R_t^*)
= ({{385+76\eta}\over{3(170+32\eta)}},~ {{10}\over{85+16\eta}}, ~
{{10+2\eta)}\over{85+16\eta}}). 
\label{6thrstarsolution} 
\end{equation}
Interestingly, these fixed point values for the couplings
$R''^*, R_b^*$ and  $R_t^*$ are same as in the minimal supersymmetric
standard model with third generation baryon number violation~\cite{anant}.

We can also try to find fixed point solutions where three of the 
trilinear couplings attain zero fixed point values, whereas the 
remaining two attain non-trivial fixed point values. In this case 
we find the following theoretically acceptable fixed points:
\begin{eqnarray}
& \displaystyle R_{\lambda}^* = R_k^* = R_b^* = 0,~~~~~ 
(R''^*,~ R_t^*)  =  ({{19}\over {24}},~  {{1}\over {8}}), &
\label{7thrstarsolution}\\
& \displaystyle R_{\lambda}^* = R_k^* = R''^* = 0,~~~~~ 
(R_b^*,~ R_t^*) = ({{35}\over {3(35+6\eta)}},~ 
{{35+7\eta}\over {3(35+6\eta)}}), &
\label{8thrstarsolution}\\
& \displaystyle R_{\lambda}^* = R_k^* = R_t^* = 0,~~~~~ 
(R_b^*,~ R''^*) = ({{2}\over {16+3\eta}},~ 
{{76+15\eta}\over {6(16+3\eta)}}). &
\label{9thrstarsolution}
\end{eqnarray}
All the infra-red fixed point solutions in (\ref{7thrstarsolution})
- (\ref{9thrstarsolution}) have one thing in common, namely
the trilinear couplings $\lambda$ and $k$ approach 
the fixed point value zero in the infra-red region. Furthermore, 
the other trilinear couplings approach 
the same infra-red fixed point values as in MSSM~\cite{anant}.

Having obtained more than one theoretically acceptable infra-red
fixed points in the NMSSM with baryon number violation, it is
important to determine which, if any, is likely 
to be realized in nature. To this end we must examine the
stability of each of the fixed point solutions (\ref{6thrstarsolution})
- (\ref{9thrstarsolution}).

We first consider the stability of the fixed point solution 
(\ref{6thrstarsolution}).  Since in this case the fixed points
of the couplings  $R_{\lambda}^* = R_k^*  = 0$, 
we have to obtain the behaviour of these
couplings around the origin.  This behaviour is determined by the
eigenvalues~\cite{allanach}
\begin{equation}\label{firstevequation}
\lambda_i={1\over b_3}\left[ \sum_{j=3}^5 S_{ij} R_j^* -(r_i+b_3)\right],
~~~ i =  1, 2,
\end{equation}
where $r_1 = r_2 = 0$,
and the matrix $S$ is the matrix 
appearing in the corresponding RG equation (\ref{componentequation}) 
but now with the ordering of the ratios as 
$R_i = (R_{\lambda}, R_k, R'',  R_b, R_t)$, 
and the fixed points
$R_i^*,\, i= 3, 4, 5$  corresponding to the fixed 
point values of $R''^*$, $R_b^*$, and $R_t^*$ 
in Eq.~(\ref{6thrstarsolution}).  
Inserting these values in Eq.~(\ref{firstevequation}), we find
\begin{equation}\label{lambda1equation}
[\lambda_i]_{i = 1, 2} = [-{{2517+ 494 \eta} \over {9(85+16\eta)}},~ -1],
\end{equation}
thereby indicating that the fixed point is attractive
in the infra-red direction.
The behaviour of the couplings $R'', R_b$ and $R_t$ around 
their respective
fixed points is governed by the 
sign  of the eigenvalues of the matrix $A$ whose
entries are ($i$ not summed over)~\cite{allanach}
\begin{equation}\label{1tildematrix}
A_{ij}={1\over b_3} R_i^* \tilde S_{ij},
\end{equation}
where $R_i^* = (R''^*, R_b^*,  R_t^*)$ 
is the fixed point solution in (\ref{6thrstarsolution}), 
and $\tilde S_{ij}$ is the upper left corner
$3 \times 3$ submatrix of the matrix (\ref{Smatrix}).
For stability, we require all the eigenvalues
of the matrix (\ref{1tildematrix}) to have negative real parts
(note that the QCD $\beta$-function $b_3 = -3$ is negative).  
The eigenvalues of the matrix (\ref{1tildematrix}) are calculated
to be 
\begin{equation}\label{lambda345}
[\lambda_i]_{i = 3, 4, 5} = [-1.6, -0.2, -0.2]
\end{equation}
which shows that the fixed point (\ref{6thrstarsolution}) 
is an infra-red stable fixed point. We note that the eigenvalue
$\lambda_3$ is larger in magnitude as compared 
to the other eigenvalues   in (\ref{lambda345}), 
indicating that the non-trivial fixed point for $\lambda_{233}^{''}$
is more attractive, and hence more relevant.

Next, we consider the stability of the fixed point solution 
(\ref{7thrstarsolution}).  Since in this case the fixed point
of the couplings
$R_{\lambda}^* = R_k^* = R_b^* = 0$,
we have to obtain the behaviour of these
coupling around the origin.  This behaviour is determined by the
eigenvalue~\cite{allanach}
\begin{equation}\label{secondevequation}
\lambda_i={1\over b_3}\left[ \sum_{j=4}^5 S_{1j} R_j^* -(r_1+b_3)\right],
~~~ i =  1, 2, 3,
\end{equation}
where $r_1 = r_2 = 0, r_3 = 2 (C_Q+C_{\overline{D}})= 16/3$,  
and the matrix $S$ is the matrix 
appearing in the corresponding RG equation (\ref{componentequation}) with
ordering of the ratios as $R_i = (R_{\lambda}, R_k, R_b, R'', R_t)$, 
and the fixed points
$R_i^*,\, i= 4, 5$  corresponding to the fixed 
point values of $R''^*$ and $R_t^*$ 
in Eq.~(\ref{7thrstarsolution}).  
Inserting these values in Eq.~(\ref{secondevequation}), we find
\begin{equation}\label{lambda2equation}
[\lambda_i]_{i = 1, 2, 3} = [ -{{9}\over{8}},~ -1,~ {{5} \over {24}}],
\end{equation}
showing thereby that the fixed point 
(\ref{7thrstarsolution}) is unstable in the infra-red region.
Similarly, it can be shown that the fixed point solutions
(\ref{8thrstarsolution}) and (\ref{9thrstarsolution})
are unstable fixed points.

One may also consider  the case where the couplings
$\lambda''_{233}, h_b, \lambda$ and $k$ attain 
trivial fixed point values, 
whereas $h_t$ attains a non-trivial fixed point value.
In this case we have $R^*_5\equiv R^*_t=7/18$, which is the same
as the Pendleton-Ross~\cite{pendross} top-quark
fixed point of the MSSM.
We must, of course, study the 
stability of this solution in the present context.
To do so, we must consider the eigenvalues
\begin{equation}
\lambda_i={1\over b_3}(S_{i5}R^*_5-(r_i+b_3)), \, i=1, 2, 3, 4, 
\label{rtfix1}
\end{equation}
where $S_{i5}$ are read off from the matrix (\ref{Smatrix}), 
but now with the couplings reorderd according to 
$R_i = (R'', R_b, R_{\lambda}, R_k, R_t)$, thereby yielding
\begin{equation}
[\lambda_i]_{i = 1,2,3,4}=[{{38}\over {27}},~ {{35}\over{54}},~ 
-{{25}\over{18}},~ -1].
\label{rtfix2}
\end{equation}
Since the sign of each of $\lambda_1$ and $\lambda_2$ is positive,
this solution is also unstable in the infra-red region.
Nevertheless, from our discussion of infra-red fixed
point solution (\ref{6thrstarsolution}),  it is clear
that the Pendelton-Ross fixed point would be
stable in the NMSSM 
in case $h_b$ and $\lambda^{''}_{233}$ are small,  though
negligible at the GUT scale (with $\lambda = k = 0$). 
In this case, these would, of course, 
evolve away from zero at the weak scale, though realistically they would 
still be small (but not zero) at the  weak scale.
Thus, the only true infra-red stable fixed
point solution is the baryon number, and  $R_p$,  violating solution 
(\ref{6thrstarsolution}). We note that this fixed point solution is
identical to the corrresponding fixed point solution in 
the MSSM with baryon number violation~\cite{anant}.
This is one of the main conclusions of this paper.
We note that the value of $R_t^*$ in (\ref{6thrstarsolution}) is lower than
the corresponding value of $7/18$ in MSSM and NMSSM
with $R_p$ conservation.

It is appropriate to examine the implications of
the value of $h_t(m_t)$ predicted by our
fixed point analysis for the top-quark mass. 
From (\ref{6thrstarsolution}),
and $\alpha_3(m_t)\simeq 0.1$,
the fixed point value for the top-Yukawa 
coupling is predicted to be $h_t(m_t) \simeq 0.4$. This translates into
a top-quark (pole) mass of about $m_t \simeq 70 \sin\beta$ GeV, 
which is incompatible with the measured value~\cite{topmass} of 
top mass, $m_t \simeq 174$ GeV, 
for any value of $\tan\beta$. 
It follows that the true fixed point obtained here
provides only a qualitative understanding of the top quark mass
in NMSSM with $R_p$ violation.

We now turn to the study of the renormalization group evolution for
the lepton number, and $R_p$,  violating couplings in the
superpotential (\ref{Lviolating}).  Here we shall consider
the dimensionless couplings $\lambda_3$, 
$\lambda_{233}$ and $\lambda'_{333}$ only. Furthermore,
we shall restrict our attention to one kind of lepton
number violation at a time. Thus, 
we shall consider three different cases,
i.e., we shall take $ \lambda_{333}^{'} \gg \lambda_3, \lambda_{233}$,
or $\lambda_{233} \gg \lambda_3, \lambda_{333}^{'}$, or $\lambda_3 \gg
\lambda_{233}, \lambda_{333}^{'}$, respectively. In the case when
$\lambda_{333}^{'}$ is the dominant of the lepton number violating
couplings, we define $R^{'} = \lambda^{'}_{333}/g_3^2$, and 
reorder the couplings as $R_i = (R^{'}, R_b, R_t,
R_{\lambda}, R_k)$, so that   
the RGEs for this case can be written as 
\begin{equation}
\frac{dR_i}{dt} = \tilde\alpha_3 R_i [(r_i+b_3)-\sum_j S_{ij}R_j],
\label{lam333}
\end{equation}
where S is the anomalous dimension matrix 
\begin{eqnarray}
S = \left[\begin{array}{ccccc}6 & 6+\eta & 1 & 0 & 0\\
6 & 6+\eta & 1 & 1 & 0 \\
1 & 1 & 6 & 1 & 0 \\
0 & 3+\eta & 3 & 4 & 2 \\
0 & 0 & 0 & 6 & 6
\end{array}\right],
\label{S333}
\end{eqnarray}
$\eta=h_\tau^2(m_t)/h_b^2(m_t)\simeq 0.4$ is the factor
coming from Eq.~(\ref{2ndhtaurelation}), and other quantities 
are defined in a manner similar to the case of baryon number violation.
The non-trivial fixed points are given by Eq.(\ref{ristarequation}), 
and are calculated to be
\begin{equation}
(R^{'*}, R_b^*, R_t^*, R_{\lambda}^*, R_k^*) = 
({{420 + 92\eta}\over{315 + 114\eta}}, 
-{{315}\over{315 + 114 \eta}}, 
{{105 + 29 \eta}\over{315 + 114 \eta}}, 
0, -2). 
\label{10thrstarsolution} 
\end{equation}
Since all the fixed point values are not positive, this is an 
unphysical solution. Thus, we conclude that a simultaneous
non-trivial fixed point for  all the couplings 
$\lambda^{'}_{333}, h_b, h_t, \lambda$ and $k$ does not exist.
We next try to find a fixed point solution with $R_b^*=0$, and
$R'^*, R_t^*, R_{\lambda}^*$ and $R_k^*$
being given by their non-zero solutions, 
relevant for low $\tan\beta$ values. 
In this case we find the fixed point solution
\begin{equation}
R_b^*=0, ~~~~(R^{'*}, R_t^*, R_{\lambda}^*, R_k^*) = 
({{10}\over{39}}, {{53}\over{78}}, -{{305}\over{156}}, {{237}\over {156}}),
\label{11thrstarsolution} 
\end{equation}
which contains a non-physical negative value, and 
hence is not an acceptable fixed point solution. 
We next try to find a
solution with two  of the trilinear couplings 
attaining a zero fixed point value. We found that 
there are no such physically acceptable fixed points. 
We then tried to find fixed points
with three of the trilinear couplings attaining a zero 
fixed point value.
There is only one such acceptable fixed point:
$R_b^* = R_{\lambda}^* = R_k^* = 0,
R^{'*} = R_t^* = {{1}\over {3}}.$
However, stability analysis shows that this fixed point
is either a saddle point, or an ultra-violet fixed point.
Furthermore, there are
no fixed points with four of the couplings attaining a
zero fixed point value. We conclude that there are no non-trivial
stable fixed points in the infra-red region for the lepton
number violating coupling $\lambda^{'}_{233}$.

If on the other hand the coupling $\lambda_{233}$ is the dominant
of the lepton number violating couplings, 
we define $R = \lambda^2_{233}/g_3^2$, and reorder the Yukawa couplings
as $(R, R_b, R_t, R_{\lambda}, R_k)$, so that the 
relevant RGEs can
be written as
\begin{equation}
\frac{dR_i}{dt} = \tilde\alpha_3 R_i [(r_i+b_3)-\sum_j S_{ij}R_j],
\label{lam233}
\end{equation}
with the anomalous dimension matrix
\begin{eqnarray}
S = \left[\begin{array}{ccccc}4 & 4\eta & 0 & 0 & 0\\
0 & 6+\eta & 1 & 1 & 0\\
0 & 1 & 6 & 1 & 0\\
0 & 3+\eta & 3 & 4 & 2\\
0 & 0 & 0 & 6 & 6
\end{array}\right],
\label{s233}
\end{eqnarray}
with the by now usual definitions of the various quantities in
(\ref{lam233}) and (\ref{s233}). The non-trivial fixed
point  is then calculated to be
\begin{equation}
(R^*, R_b^*, R_t^*, R_{\lambda}^*, R_k^*) = 
(-{{90 + 109\eta}\over{12(10 + \eta)}}, 
{{25}\over{3(10 + \eta)}}, 
{{5(5 + \eta)}\over{3(10 + \eta)}}, 
-{{105 + 23\eta}\over{3(10 + \eta)}}, 
{{30 + 43\eta}\over{6(10 + \eta)}}), 
\label{12thrstarsolution} 
\end{equation}
which must be rejected as unphysical. We next try to find a fixed point
solution with $R_b^*=0$, and
$R'^*, R_t^*, R_{\lambda}^*$ and $R_k^*$
being given by their non-zero solutions. In this case we find the fixed 
point solution
\begin{equation}
R_b^* = 0, ~~~~
(R^*, R_t^*, R_{\lambda}^*, R_k^*) = 
(-{{3}\over{4}}, {{20}\over{27}}, -{{19}\over{9}}, {{29}\over {18}}),
\label{13thrstarsolution} 
\end{equation}
which, again,  is physically unacceptable. 
Proceeding further as in the previous
cases, we have not found any physically acceptable 
stable infra-red fixed points
in this case. Thus, there are no fixed point solutions for the 
lepton number violating coupling $\lambda_{233}.$

Finally, if the lepton number violating coupling $\lambda_3$ is the
dominant coupling, we define $R_3 = \lambda_3^2/g_3^2$, 
and reorder the couplings as $(R_3, R_b, R_t,
R_{\lambda}, R_{\kappa})$, so that the RGEs can be written in the form
\begin{equation}
\frac{dR_i}{dt} = \tilde\alpha_3 R_i [(r_i+b_3)-\sum_j S_{ij}R_j],
\label{lam3}
\end{equation}
with the anomalous dimension matrix
\begin{eqnarray}
S = \left[\begin{array}{ccccc}4 & \eta & 3 & 4 & 2\\
0 & 6+\eta & 1 & 1 & 0\\
1 & 1 & 6 & 1 & 0\\
4 & 3+\eta & 3 & 4 & 2\\
6 & 0 & 0 & 6 & 6
\end{array}\right],
\label{s3}
\end{eqnarray}
which leads to the fixed point values for the couplings
\begin{equation}
(R^*_3, R_b^*, R_t^*, R_{\lambda}^*, R_k^*) = 
(-{{100}\over{27}}, 0, 
{{20}\over{27}}, 
{{43}\over{27}}, 
{{29}\over{18}}).
\label{14thrstarsolution} 
\end{equation}
This fixed point is physically unacceptable. Thus, in this case
also there is no  fixed point with all  couplings approaching
non-trivial fixed point values.

We can next try a fixed point with $R_b^* = 0$, which is relevant
for low values of $\tan\beta.$ In this case we 
find the fixed point
\begin{equation}
R_b^* = 0, ~~~~~
(R_3^*, R_t^*, R_{\lambda}^*, R_k^*) = 
(0, {{20}\over{27}}, -{{19}\over{9}}, {{29}\over {18}}),
\label{15thrstarsolution} 
\end{equation}
which is again unacceptable. We now try to obtain fixed point 
solution with two of the couplings
$R_3^* = R_b^* = 0$, and other couplings having 
non-trivial fixed point values. We find
\begin{equation}
R_3^* = R_b^* = 0, ~~~~~ (R_t^*, R_{\lambda}^*, R_k^*) = 
({{20}\over{27}},
-{{19}\over{9}}, 
{{29}\over{18}}),
\label{16thrstarsolution} 
\end{equation}
which must also be rejected. We have also tried to obtain a fixed point
solution with $R_{\lambda}^* = R_k^* = 0$,  with other couplings having
non-zero fixed point values. We find
\begin{equation}
R_{\lambda}^* = R_k^* = 0, ~~~~~ (R_3^*, R_b^*, R_t^*) = 
(-{{210+55\eta}\over{3(61+11\eta)}},
{{55}\over{3(61+11\eta)}}, 
{{97+22\eta}\over{3(61+11\eta)}}),
\label{17thrstarsolution} 
\end{equation}
which, again, is unacceptable. We have also checked that the
trivial fixed point for the couplings $\lambda_3, h_b, \lambda, k$
and the Peddelton-Ross type fixed point for the top-quark Yukawa coupling
is unstable in the infra-red region. We, therefore, conclude that
in this case of lepton number 
violation also there are no acceptable  infra-red  fixed points.

Thus, there are no non-trivial, infra-red stable fixed points
for any of the lepton number violating couplings in the
non-minimal supersymmetric standard model.

To conclude, we have  analyzed the one-loop
renormalization group equations for the evolution of Yukawa couplings
in the non-minimal supersymmetric standard model 
with $R_p$ violating couplings 
to the heaviest generation, taking into account 
B and L violating couplings one at a time.
The analysis of model 
yields the surprising and important result that  
only the simultaneous non-trivial
fixed point for the baryon number violating coupling 
$\lambda^{''}_{233}$  and the top-quark and b-quark
Yukawa couplings $h_t$ and $h_b$, and the  
trivial fixed point for $\lambda$ and $k$,
is stable in the infra-red region.
However, the fixed point value for the top-quark coupling here is 
lower than its corresponding value in the MSSM, and NMSSM, with 
R-parity conservation,
and is incompatible with the measured value of the top-quark mass. 
Thus, it appears that the baryon number, and $R_p$, violating 
coupling has the effect of reducing the infra-red fixed point value
of the top-quark Yukawa coupling to the same extent in MSSM and NMSSM.
The $R_p$ conserving solution with $\lambda''_{233}$ attaining its
trivial fixed point, with $h_t$ and $h_b$ attaining non-trivial
fixed points, is infra-red unstable, as is the case for 
trivial fixed points for $\lambda''_{233}$ and $h_b$, with
a non-trivial fixed point for $h_t$.  
Our analysis
shows that the usual Pendleton-Ross
type of infra-red fixed point is unstable in the presence of 
$R_p$ violation, though for small, but negligible,  values of 
$h_b$ and $\lambda^{''}_{233}$ it could be stable.
We have also found that there are no
non-trivial infra-red stable fixed points
for the lepton-number, and R-parity, violating
couplings in the NMSSM. Our results
are the first in placing strong theoretical constraints on the nature
of $R_p$ violating couplings in the NMSSM from fixed-point and stability
considerations: the fixed points that are unstable, or the 
fixed point that is a saddle point, cannot be realized in 
the infra-red region. 
The fixed points obtained in this work
are the true fixed points, and serve
a lower bound on the relevant $R_p$ violating Yukawa couplings.
Their structure is essentially the  same as in MSSM with
R-parity violation.
In particular, from our analysis of the simultaneous (stable) fixed
point for the baryon number violating coupling $\lambda^{''}_{233}$
and the top and bottom Yukawa couplings, we infer a lower bound
on $\lambda^{''}_{233} \stackrel{>}{\sim} 0.98$.

\begin{small}
\noindent {\bf Acknowledgements:}
The work of  (PNP) is supported by the 
University Grants Commission research award
No. 30-63/98/SA-III.
\end{small}


\newpage

\begin{thebibliography}{abcdef}

\bibitem{pendross} B. Pendleton and G. G. Ross, Phys. Lett. {\bf B98}
(1981) 291; C. T. Hill, Phys. Rev. {\bf D 24} (1981) 691;
M. Lanzagorta and G. G. Ross, Phys. Lett. {\bf B349} (1995) 319.

\bibitem{schrempp} See, e.g., B. Schrempp and M. Wimmer, 
Prog. Part. Nucl. Phys. {\bf 37}, 1 (1996).

\bibitem{nilles} H. P. Nilles, Phys. Rep. {\bf 110} (1984) 1;
H.E. Haber and G.L.  Kane, {\it ibid} {\bf 117} (1985) 75.

\bibitem{weinberg} S. Weinberg, Phys. Rev. {\bf D 26} (1982) 287;
N. Sakai and T. Yanagida, Nucl. Phys. {\bf B 197} (1982) 133.

\bibitem{farrar} G. Farrar and P. Fayet, Phys. Lett. {\bf B76} (1978) 575.

\bibitem{barger} V. Barger et al.,
Phys. Rev. {\bf D 53} (1996) 6407; 
B. C. Allanach, A. Dedes and H. K. Dreiner, hep-ph/9902251, 
and references therein.

\bibitem{anant} B. Anathanarayan and P. N. Pandita, 
Phys. Lett. {\bf B454} (1999) 84.

\bibitem{allanach} B. C. Allanach and S. F. King, Phys. Lett. 
{\bf B407} (1997) 124;
S. A. Abel and B. C. Allanach, Phys. Lett. {\bf B415}
(1997) 371.

\bibitem{fayet} P. Fayet, Nucl. Phys. {\bf B90} (1975)104;
R. K. Kaul and P. Majumdar, Nucl. Phys. {\bf B199} (1982)36;
J. Ellis et al., Phys. Rev. {\bf D39} (1989) 844. 

\bibitem{dreiner} See, e.g., B. C. Allanach, A Dedes and H. K.  Dreiner, 
hep-ph/9906209.

\bibitem{smirnov} A. Y. Smirnov and F. Vissani, Phys. Lett. 
{\bf B380} (1996) 317.

\bibitem{barger2} V. Barger, M. S. Berger and P. Ohmann,
Phys. Rev. {\bf D 47} (1993) 1093.

\bibitem{topmass} Particle Data Group, C. Caso et al., Eur. Phys. J. 
C3 (1998) 1.

\end{thebibliography}
\end{document}